\documentclass{article}

\usepackage{latexsym}

\begin{document}

\newtheorem{theorem}{Theorem}
\newtheorem{proposition}{Proposition}
\newtheorem{remark}{Remark}
\newtheorem{corollary}{Corollary}
\newtheorem{lemma}{Lemma}
\newtheorem{observation}{Observation}
\newtheorem{definition}{Definition}

\newcommand{\qed}{\hfill$\Box$\medskip}

\title{A Note on Kolmogorov-Uspensky Machines} 
\author{Holger Petersen\\
Reinsburgstr. 75\\
70197 Stuttgart\\
Germany 
}

\maketitle

\begin{abstract}
Solving an open problem stated by Shvachko, it is shown that a language
which is not real-time recognizable by some variants of pointer machines
can be accepted by a Kolmogorov-Uspensky machine in real-time.
\end{abstract}

\section{Introduction}
The Kolmogorov-Uspensky machine (KUM) is a very general model of
sequential computation that was introduced in 1953.  
The article \cite{KU63} gives a detailed
description and shows that all recursive functions are computable by
KUMs. A closely related model that was independently introduced by
A.~Sch{\"o}nhage \cite{Schonhage70,Schonhage80} is the storage
modification machine (SMM). While the KUM works on an undirected graph
with bounded degree, the SMM is equipped with a directed graph of
bounded out-degree but possibly unbounded in-degree.

Real-time computation will be understood in the sense of \cite{Gurevich88}.
The number of steps carried out by a machine between reading and writing 
successive symbols is bounded by a constant when working in real-time. 
This notion of real-time computation is preserved by a simulation between 
different classes of machines, if it satisfies the following definition
due to Sch{\"o}nhage \cite{Schonhage80}:
\begin{definition}
A machine $M'$ is said to simulate in real-time another machine $M$,
if there exists a constant $c$ such that for every input $x$ the following holds: if
$x$ causes $M$ to read an input symbol, or to print an output symbol, or to halt at time
steps $0 = t_0 < t_1 < \cdots < t_{\ell}$, respectively, then $x$ 
will cause $M'$ to act in the same
way with regard to those external instructions at time steps 
$0 = t_0' < t_1' < \cdots < t_{\ell}'$, where
$t_i'-t_{i-1}' \le c(t_i-t_{i-1})$ for $1\le i\le \ell$.
\end{definition}

Shvachko \cite{Shvachko91} defined the language
$$L = \bigcup_n\{b_{0^n}@\ldots @b_{1^n}\# x\# y\#\mid 
b_i\in\{0,1\}^{[n/2]}, x, y\in\{0, 1\}^n, b_x = b_y\}$$ 
and showed that there is an SMM recognizing $L$ in real-time.
The main idea is to represent every binary string $i$ of length
$n$ as a path in a tree. From the leaf reached in the tree
an edge points to a vertex that represents $b_i$.
Equality of $b_i$'s can then be tested in constant time. Since 
there is no upper bound on the number of $i$'s with
equal $b_i$ this graph in general has unbounded in-degree.
Shvachko established that three variants of pointer machines
cannot accept $L$ in real-time, but left open if KUMs 
are powerful enough to solve the problem in real-time.
A negative answer would solve the long standing open 
problem about the possibility of a real-time simulation 
of SMMs by KUMs, since a real-time simulation according to the 
above definition together with the algorithm from \cite{Shvachko91}
for a SMM would yield a corresponding solution for a KUM.

\section{The Result}
In this section we present a real-time algorithm for language $L$ defined
above, that can be carried out by a KUM. This solves the open problem 
from \cite{Shvachko91}.
\begin{theorem}
There is a KUM that accepts $L$ in real-time.
\end{theorem}
{\bf Proof.} The KUM $M$ accepting $L$ keeps two trees and a number of
auxiliary data structures while reading an input. The first tree $A$
is a complete binary tree up to level $n$ built by $M$ while reading
the portion of the input until the first \#.  Each node at level $n$
represents an $i$ and after forming the path to this node, the KUM
attaches a newly created string encoding $b_i$ to this node.
Concurrently with building tree $A$ the machine forms a binary tree
$B$ such that nodes at level $[n/2]$ represent the $b_i$.  Beyond this
level a possibly incomplete binary tree of depth $n$ is formed. A path
representing $i$ is formed in this tree if $b_i$ matches the value
represented in the upper portion of $B$. Here we have to overcome a
problem: The node in the upper tree is only determined if all of $b_i$
has been read. Therefore the construction of the path in the lower
part is done while reading $b_{i+1}$ at twice the speed of reading
bits from the input. For $b_{1^n}$ the construction is done while
reading $x$.

We now describe the algorithm carried out by $M$ in more detail. 
The computation on the base segment $B = b_{0^n}@\ldots @b_{1^n}$
is split into $2^n$ phases, where in phase $i$ the string $b_i$ is processed.
For $1 \le i < 2^n$ each phase consists of the following activities, which
are carried out in an interleaved fashion in order to obtain a real-time
solution:
\begin{itemize}
\item A counter consisting of $n$ bit positions is incremented from $i-1$
(the value from the previous phase) to $i$ handling two bits for each symbol read
from the input. 
If the symbol $@$ is reached before or after the counter is completely processed,
the input is rejected.
\item A path representing $i-1$ is constructed in $B$ starting at the leaf corresponding
to $b_{i-1}$, creating new nodes if necessary. Two levels are constructed for 
each symbol read from the input. 
\item A path for $i$ is constructed in $A$, creating new nodes if necessary. Again 
two levels are processed for each symbol read from the input. 
\item A path for $b_i$ is constructed in $B$.
\item A string representing $b_i$ is constructed.
\end{itemize}
Phase $0$ deviates from the description above, since the counter has to be initialized
and no path representing $i-1$ is constructed in $B$.

At the end of phase $i$ the leaf constructed in $A$ is linked to the string 
representing $b_i$. 

If $M$ reaches $x$, it checks that the counter has reached the value $2^n-1$ 
(this can be determined while incrementing the value).  
Then $M$ traverses the path corresponding to $x$ in
tree $A$ and concurrently constructs the missing path representing $2^n-1$
starting at the leaf for $b_{2^n-1}$. 

The computation on $y$ is split into two phases. In the
first phase $M$ traverses the path corresponding to $b_x$ (linked to the
leaf of $x$ in $A$) in the tree $B$.  While doing so, $M$ stores the first half
of $y$ on a queue.  In the second phase it checks whether $y$ is
stored in the portion of $B$ encoding all $i$ with $b_i = b_x$. 
To this end $M$ processes two bits
from the queue while reading one input bit (which is appended
to the queue).  It is therefore possible
to reach a leaf of $B$ when the input has been completely
consumed.\qed

\section{Discussion}
We have shown that the language $L$ from \cite{Shvachko91} can be accepted by a
KUM in real-time. It is essential for the approach presented that
reading the ``address'' $y\in\{0, 1\}^n$ leaves enough time for preparing
the equality check. A natural modification, for which we do not have
a real-time algorithm, would be to shuffle the bits of $x$ and $y$. It remains open
whether this modified language can separate KUM and SMM in real-time.


\begin{thebibliography}{Gur88}

\bibitem[Gur88]{Gurevich88}
Yuri Gurevich.
\newblock The logic in computer science column.
\newblock {\em Bulletin of the EATCS}, 35:71--82, 1988.

\bibitem[KU63]{KU63}
A.~N. Kolmogorov and V.~A. Uspensky.
\newblock On the definition of an algorithm.
\newblock {\em American Mathematical Society Translations}, 29:216--245, 1963.
\newblock Translation of Uspekhi Mat. Nauk 13:3--28, 1958.

\bibitem[Sch70]{Schonhage70}
A.~Sch{\"o}nhage.
\newblock Universelle {T}uring {S}peicherung.
\newblock In J.~D{\"o}rr and G.~Hotz, editors, {\em Automatentheorie und
  formale Sprachen}, volume~3 of {\em Berichte aus dem mathematischen
  Forschungsinstitut Oberwolfach}, pages 369--383, Mannheim, 1970.
  Bibliographishes Institut.

\bibitem[Sch80]{Schonhage80}
A.~Sch{\"o}nhage.
\newblock Storage modification machines.
\newblock {\em SIAM Journal on Computing}, 9:490--508, 1980.

\bibitem[Shv91]{Shvachko91}
Konstantin~V. Shvachko.
\newblock Different modifications of pointer machines and their computational
  power.
\newblock In Andrzej Tarlecki, editor, {\em Proceedings of Mathematical
  Foundations of Computer Science. ({MFCS} '91)}, volume 520 of {\em LNCS},
  pages 426--435, Berlin-Heidelberg-New York, 1991. Springer.

\end{thebibliography}
%

\end{document}